\documentclass[11pt, journal,draftclsnofoot,onecolumn]{IEEEtran}
\usepackage{microtype}
\usepackage{amsmath,amsfonts}
\usepackage{algorithmic}
\usepackage{arydshln}
\usepackage{array}
\usepackage{multirow}
\usepackage{diagbox}
\usepackage{caption}
\usepackage{textcomp}
\usepackage{stfloats}
\usepackage{threeparttable}
\usepackage[T1]{fontenc}
\usepackage[utf8]{inputenc}
\usepackage{babel}
\usepackage[font=small,labelfont=bf,tableposition=top]{caption}
\usepackage{booktabs}
\usepackage{threeparttable} 
\usepackage{setspace} 
\usepackage{diagbox}
 \usepackage{bigstrut}
\usepackage{tabularx}\usepackage{url}
\usepackage{dashrule}
\usepackage{booktabs}
\usepackage{verbatim}
\usepackage{graphicx}
\usepackage{tabularx} 
\usepackage{arydshln}
\def\BibTeX{{\rm B\kern-.05em{\sc i\kern-.025em b}\kern-.08em
    T\kern-.1667em\lower.7ex\hbox{E}\kern-.125emX}}
\usepackage{balance}
\usepackage{microtype}
\usepackage{titlesec}
\titlespacing{\subsection}{0pt}{\parskip}{-\parskip}
\begin{document}
\title{ An Efficient Network with Novel Quantization Designed for Massive MIMO CSI Feedback}
\author{Xinran Sun,
        Zhengming Zhang,~\IEEEmembership{Student Member,~IEEE,}
        Luxi Yang,~\IEEEmembership{Senior Member,~IEEE}}

\maketitle
\begin{abstract}
\fontsize{11}{12}\selectfont
The efficacy of massive multiple-input multiple-output (MIMO) techniques heavily relies on the accuracy of channel state information (CSI) in frequency division duplexing (FDD) systems. Many works focus on CSI compression and quantization methods to enhance CSI reconstruction accuracy  with lower feedback overhead. In this letter, we propose CsiConformer, a novel CSI feedback network that combines convolutional operations and self-attention mechanisms to improve CSI feedback accuracy. Additionally, a new quantization module is developed to improve  encoding efficiency. Experiment results show that CsiConformer outperforms previous state-of-the-art networks, achieving an average accuracy improvement of 17.67\% with lower computational overhead. 
\end{abstract}


\section{Introduction}
\IEEEPARstart{M}{ASSIVE} multi-input multi-output (MIMO) is one of the most prominent technologies in current 5G and future 6G communication systems. With advanced techniques like spatial multiplexing and beamforming, a 5G base station (BS) with  multiple antenna arrays can effectively suppress multiuser interference and deliver high spectrum \cite{hong2017multibeam}.

The accuracy of the downlink channel state information (CSI) directly affects the performance of massive MIMO systems. In time-division duplex (TDD) systems, channel reciprocity is beneficial to estimation of downlink CSI, as the fading characteristics between downlink and uplink channels exhibit minimal differences \cite{sanguinetti2019toward}. In contrast, in frequency division duplex (FDD) systems, reciprocity between the uplink and downlink channels is absent. Downlink CSI is first estimated at the user equipment (UE) using pilots from the BS.  It is then transmitted back to the BS using feedback link. However, with the increasing number of antennas, receivers and subcarriers, the substantial uplink bandwidth overhead required for downlink CSI feedback becomes  unacceptable for effective deployment of massive MIMO systems. It is necessary to efficiently compress and quantize downlink CSI before transmission.

 CSI compression techniques mainly include compressive sensing (CS) and deep learning (DL)-based methods. Traditional CS methods \cite{kuo2012compressive}, relying on the assumption of  channel sparsity, are hindered by their efficiency in iteratively reconstructing \cite{kyritsi2003correlation}.  DL-based methods have surpassed traditional CS methods and shown promising results in downlink CSI feedback by treating a CSI matrix as an image. The first DL-based framework, CsiNet \cite{wen2018deep}, utilizes  convolutional neural networks (CNNs) and significantly outperforms CS methods. Since then, further advancements have extended the original scene, including channel implicit reciprocity \cite{liu2019exploiting}, time correlation using LSTM \cite{wang2018deep}, spatial correlation among multi-users \cite{9839111}, etc., while more researchers focus on improving the autoencoder architecture to enhance CSI feedback accuracy.  Among them, \cite{csinetjia}  increases the receptive field and extracts deeper channel features to achieve performance improvement. Notably, CRNet  \cite{CRNet} and CLNet \cite{CLnet} respectively adopt 
 convolution factorization and 1×1 point-wise
convolution to save computational resources without distinct performance decrease.  TransNet \cite{transnet} adopts a two-layer Transformer architecture to improve CSI reconstruction accuracy. DCRNet employs a dilated CNN-based autoencoder to efficiently extract features from block-sparse CSI matrices  \cite{dilated}. In \cite{hu2023learnable}, the regularization term is treated as a learnable function to better fit the CSI characteristics. Additionally, quantizing  float-point codewords is indispensable for further improving the encoding efficiency in practice. The residual learning \cite{csinetjia}, clustering schemes \cite{clusterQUA}, and approximate gradient functions \cite{changeable_rate_and_quafunction} are introduced to generate bitstream and reduce quantization distortion.

DL-based networks can be generally divided into CNNs and attention frameworks. Note that convolution operations are proficient in extracting local features, whereas the self-attention mechanism specializes in capturing global representations \cite{conformer}.  Specifically, the Transformer-based TransNet may be less effective in extracting fine-grained local details in CSI. This motivated our work to integrate these two distinct features to improve CSI compression feedback performance.

In this letter, we propose a novel network architecture, CsiConformer, which adopts a four-layer Conformer \cite{conformer} structure to efficiently integrate local features and global representations with less computational resources. Besides, we develop a new quantization module based on VQ-VAE \cite{vqvae} sandwiched between two convolution layers. Ultimately, CsiConformer outperforms the state-of-the-art (SOTA) networks, achieving 17.67\% higher accuracy and 37.36\% lower complexity on average.

The main contributions are summarized as follows:
\begin{itemize}
\item{CsiConformer is composed of four sequential Conformer layers, combining convolution and Transformer to model both local and global features of CSI in a parameter-efficient way. Comparative experiments with previous SOTA DL-based networks demonstrate that CsiConformer improves reconstruction quality of downlink CSI feedback with low computational complexity.}
\item{The proposed sandwiched VQ-VAE (SVQ-VAE) adopts an up-channel convolution to efficiently  transform CSI feature maps into a latent representation. The representation well matches with the embedding vectors in the codebook, mitigating information loss caused by CSI compression. }
\end{itemize}

The rest of this letter is organized as follows. Section II describes the system model and section III explains the detailed design of CsiConformer and SVQ-VAE. Section IV shows the training scheme and results of the proposed architecture. The conclusion is provided in Section V.
\begin{figure}[!t]
\centering
\captionsetup[figure]{name={Fig.},labelsep=period,singlelinecheck=off} 
\includegraphics[width=6.5in]{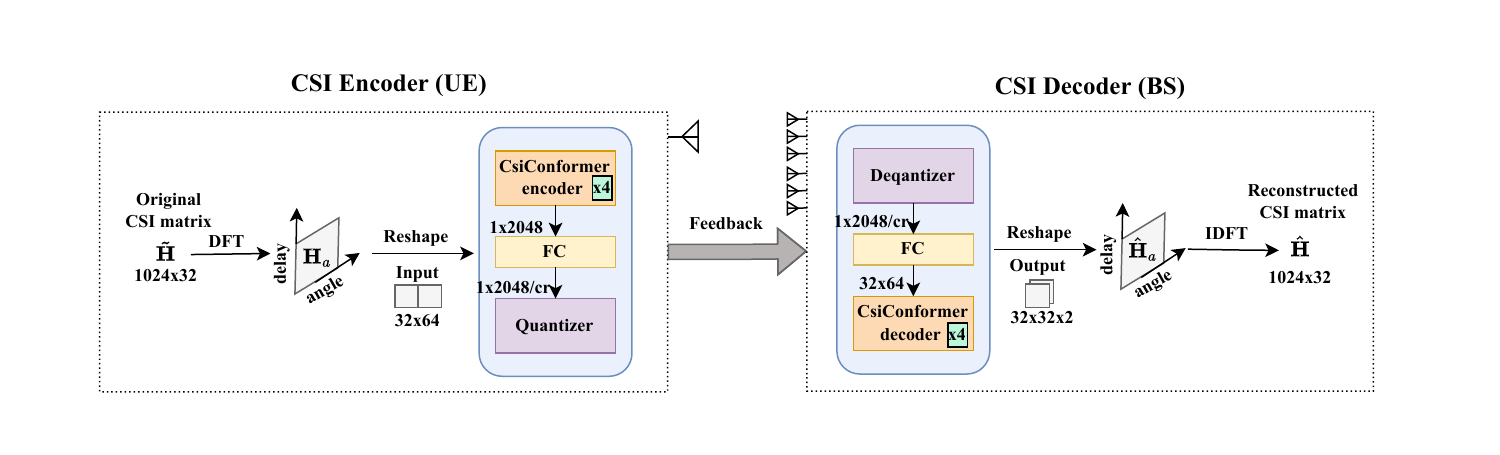}
\caption{Proposed CsiConformer framework}
\label{fig1}
\vspace{-15pt}
\end{figure}
\section{System Model}
Consider a single-cell massive MIMO FDD system with \(N_{t}\) transmitting antennas at the BS and \(N_{r}\) receiving antennas at the UE (where \( N_{t} \gg  N_ {r}\), and following \cite{wen2018deep}, \(N_{r}\) is set to 1). The CSI feedback is transmitted  from the UE to the BS with \(N_{c}\) sub-carriers. 
The CSI matrix of downlink channels in the spatial-frequency domain can be denoted as \(\tilde{{\rm \textbf{H}}} = [\tilde{{\rm \textbf{h}}}_{1},\cdots,\tilde{{\rm \textbf{h}}}_{N_{c}}]^H \in \mathbb{C}^{{N_{c}} \times{ N_{t}}}\), where \(\tilde{{\rm \textbf{h}}}_{n} \in \mathbb{C}^{N_{t} \times 1}, n\in\left\{1, \cdots, N_{c}\right\} \) represents the downlink channel vector of the \(n\)-th subcarrier. To reduce feedback overhead, the sparsity of CSI in the angular-delay domain \cite{wen2018deep} is adopted using a 2D discrete Fourier transform (2D-DFT) as 
\begin{equation}
\label{eq1}
 {\rm \textbf{H}}' = {\rm \textbf{F}}_{c} {\rm \tilde{\textbf{H}}}{\rm \textbf{F}}_{t}^{H} \text{,}
\end{equation}
where \({\rm \textbf{F}}_{c}\) and \({\rm \textbf{F}}_{t}\) are the DFT matrices with dimension  \( N_{c} \times N_{c}  \) and \( N_{t} \times N_{t}  \), respectively. Only the truncated  \( \textbf{H}_{a} \), the first \(N_{a}\) rows of \({\rm \textbf{H}}'\), contains distinct non-zero values, as the time delay among multiple paths lies within a limited period.  \( \textbf{H}_{a} \) first enters the encoder at the UE to undergo compression  operation to generate the codeword \( \textbf{H}_{c} \), which can be expressed as
\begin{equation}
\label{eq2}
{\rm {\textbf{H}}}_c = \mathcal{E}({\textbf{H}}_a , \Theta_{\mathcal{E}} ) \text{,} 
\end{equation}
\noindent where  \(\mathcal{E}(\cdot ) \)  denotes the compression process in the encoder, \( \Theta _\mathcal{E} \) represents the compression parameters. Then, the quantization module at the UE discretizes the codeword and generate the bitstream. Once the BS receives the bitstream, dequantization and decompression operations are applied to reconstruct the CSI matrix. The reconstructed codeword \({\hat{\textbf{H}}}_{c} \)  can be denoted by
\begin{equation}
\label{eq3}
{\rm \hat {\textbf{H}}}_{c} = \mathcal{DQ}(\mathcal{Q}( {\textbf{H}}_c , \Theta_{\mathcal{Q}} ))\text{,}   
\end{equation}
\noindent where \(\mathcal{Q}(\cdot ) \) and  \(\mathcal{DQ}(\cdot ) \) are the quantization and dequantization operations, respectively,  \( \Theta _\mathcal{Q} \) represents the parameters in the codebook of SVQ-VAE. Finally, the reconstructed downlink CSI matrix \(\hat{\textbf{H}}_a \) can be expressed as
\begin{equation}
\label{eq4}
{\rm \hat {\textbf{H}}}_a = \mathcal{DE}(\hat {\textbf{H}}_{c} , \Theta_{\mathcal{DE}} )\text{,}
\end{equation}
\noindent where \(\mathcal{DE}(\cdot ) \) denotes the decompression process with parameters \( \Theta _\mathcal{DE} \).
By combining equations (2)-(4), the entire feedback process is formulated as an optimization form with mean-squared error (MSE) distortion metric, given by
\begin{equation}
\label{eq5}
\mathrm{MSE}=\frac{1}{N} \sum_{n=1}^{N}\left\|\mathbf{H}_{a}^{n}-\hat{\mathbf{H}}_{a}^{n}\right\|_{F}^{2}\text{,}
\end{equation}
    \noindent where \(N\)  is the total number of training samples for establishing CSI compression feedback module, \(n\) denotes the index. The aim is to minimize the MSE by training  \({\Theta}_{\mathcal{E}}\), \({\Theta}_{\mathcal{DE}}\), and \({\Theta}_{\mathcal{Q}}\).
\section{Design of CsiConformer and SVQ-VAE}
In this section, we describe the proposed  CsiConformer and quantization module SVQ-VAE. 
\subsection{CsiConformer Design}
As shown in Fig. 1, the left UE part compresses and then quantizes the CSI before the right BS part reconstructs the CSI. In the encoder, the input of CsiConformer, sized \(N_{a}\!\times\!2N_{t}\), is a real-valued CSI matrix formed by concatenating the real and imaginary parts of \( {\textbf{H}}_a\). The reformed matrix first enters four sequential Conformer layers, with each layer's output maintaining the same size as the input. After that, the matrix is converted into a vector of length \(2N_{a}N_{t}\) before being fed into a fully-connected (FC) layer compressing the CSI features  according to a specific compression ratio (CR). In the decoder, the reshaped codeword enters a FC layer and a four-layer CsiConformer structure, resulting in the desired CSI feedback system.
\begin{figure}[!t]
\centering
\includegraphics[width=3.4in]{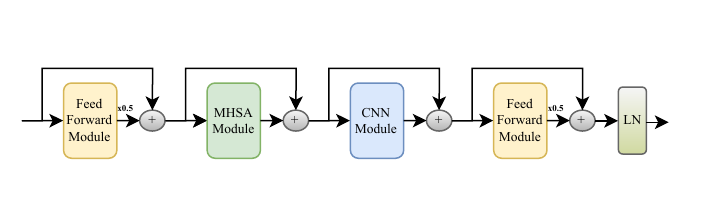}
\caption{Overview of the procedure in a Conformer layer.}
\label{fig2}
\vspace{-15pt}
\end{figure}
As depicted in Fig. 2, a conformer layer \cite{conformer} contains four modules stacked together: a feed-forward module, a multi-headed self-attention (MHSA) module without position encoding, a convolution module, and a second feed-forward module. The MHSA and convolution modules are sandwiched between two feed-forward modules with pre-norm half-step residual connections-one preceding the MHSA and the other following the convolution. The feed-forward module applies layer normalization before expanding the CSI matrix's rows using an expansion factor of 4 by the first linear layer. Subsequently, Swish activation and dropout are applied for regularization. Finally,  the second linear layer reduces the matrix's rows to the original dimension. For the CSI matrix, the MHSA module captures long-distance feature dependencies with  attention score matrices \cite{transnet} and the number of attention heads is set to 8. The convolution module primarily contains a pointwise convolution layer and a single 1D depthwise convolution layer to extract local details. Therefore,  both local and global features are maximally retained with the combination of the MHSA module and convolution module. The expansion factor of the pointwise convolution and the  kernel size of the 1D depthwise convolution is set to 2 and 31, respectively. 

\subsection{SVQ-VAE Design}
\begin{figure}[!t]
\centering
\includegraphics[width=3.5in]{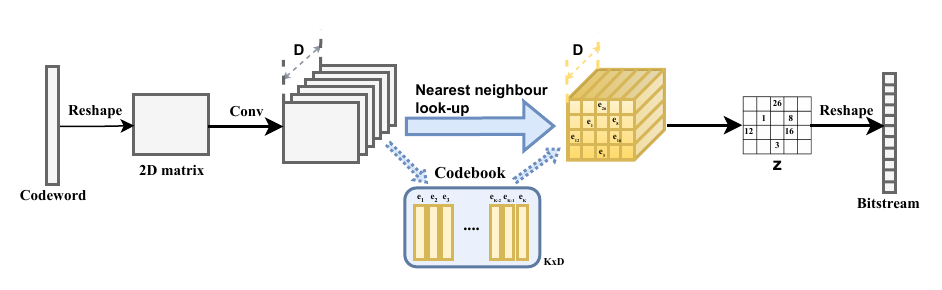}
\caption{Structure of SVQ-VAE quantizer. SVQ-VAE dequantizer is equal to the mirrored version of the quantizer.}
\label{fig3}
\vspace{-15pt}
\end{figure}
VQ-VAE learns discrete representations utilizing  a codebook in a data-driven  way, which transfers the  input into an efficient vector with specific discrete latent variables. The codebook is a latent embedding space \(e\!\in\!\mathbb{R} ^{K\times D}\)  including \(K\) latent embedding vectors \( e^{i}\), \(i\!\in\!1,2,\dots K\). In 
 the previous VQ-VAE-based CSI quantization \cite{rizzello2022learning}, a latent embedding vector incorporates adjacent elements in CSI, leading to a loss of feedback accuracy. In contrast, SVQ-VAE associates each element of a codeword with a related D-length latent vector. The detailed procedure of SVQ-VAE is illustrated in Fig. 3. Initially,  a 2D matrix, reshaped from a codeword,  increases the  channel dimension up to the length of latent embedding vectors using an up-channel convolution with a kernel size of \(1\times 1\), thereby expanding the network's capacity to learn complex and abstract features. Each element in the original codeword becomes a vector of length \(D\). The discrete latent variables \(z\) are calculated through nearest neighbor look-up to match every D-length vector with the index of a latent embedding vector in the codebook. Generally, the quantizer transforms a codeword consisting of 32-bit floating-point elements into the bitstream with particular indexes that maps every element of a codeword to 1-of-\(K\) latent embedding vectors. Subsequently, the bitstream is  transformed to the form containing latent embedding vectors and restored to the codeword of length \(2N_{a}N_{t}/\)CR using a down-channel convolution at the BS side. The training objective of SVQ-VAE becomes

\begin{small}
\begin{equation}
\label{eq6}
L_{\mathcal{Q}}\!=\!\log p\left(x\! \mid\! q_{r}(x)\right)\!+\!\left\|\operatorname{sg}\left[q_{s}(x)\right]\!-\!e\right\|_{2}^{2}\!+\!\beta\left\|q_{s}(x)\!-\!\operatorname{sg}[e]\right\|_{2}^{2}\text{,}
\end{equation}
\end{small}

\noindent where \(x\) denotes the input of the encoder, \(q_{s}(x)\) denotes the output of the  up-channel convolution, \(q_{r}(x)\) denotes the input of the down-channel convolution, \(\operatorname{sg}\) denotes the stopgradient operator, and \(\beta =0.25\) is the loss weight. The three terms are, respectively, the reconstruction loss for two convolution layers, the \(l_{2}\) error for the embedding vectors \(e_{i}\), and the commitment loss for training stabilization. The quantization strategy  enables the CSI feedback network including encoder and decoder to be trained with a differentiable function in an end-to-end fashion, tackling the back-propagation training problem. The loss function to train \({\Theta}_{\mathcal{E}}\), \({\Theta}_{\mathcal{DE}}\), and \({\Theta}_{\mathcal{Q}}\)  is given by
\begin{equation}
\label{eq7}
L_{sum}=\|{\rm \hat {\textbf{H}}}_a - {\rm  {\textbf{H}}}_a\|_{F}^{2} + L_{\mathcal{Q}}\text{.}
\end{equation}

\section{Simulations}
This section presents detailed experiment settings and compares the proposed CSI feedback network with the previous SOTA  DL-based networks in terms of accuracy and computational overhead.

\subsection{Training Scheme and Evaluation Metric}
To ensure a fair performance comparison, we exploit the widely-used dataset COST2100 \cite{cost2100} in both indoor and outdoor scenarios at the frequency of 5.3 GHz and 300 MHz, respectively. Following \cite{wen2018deep}, we set the number of antennas to \(N_t=32\) and the number of subcarriers to \(N_c=1024\) for FDD systems, while \(N_a= 32\) in the angular domain. As  shown in Fig. 1, original CSI matrices are converted to the angular-delay domain  by 2D-DFT and the two channels of \(\textbf{H}_a\in\mathbb{R} ^{N_a\times\!N_t\times2}\) represent the real and imaginary parts, respectively. \(\textbf{H}_a\) is then transferred to a real-valued matrix  with the dimension of \(32\times64\) by concatenating the two channels  before  entering the CsiConformer encoder. The second dimension actually  corresponds to the row vectors of CSI matrices in the MHSA and FC modules of the encoder and decoder, which is set to 64 considering that CsiConformer should learn CSI features with a sufficient but not excessive number of parameters. \(\mathbf{\hat{H}}_a\) can be restored by zero filling and IDFT at the BS side.

\begin{table*}[htbp]
  \centering

\small 
     \captionsetup{justification=centering}
   \caption{\centering{NMSE(dB) and FLOPs Comparision Between CsiConformer and Previous CSI Feedback Methods}}
   \begin{threeparttable} 
  \begingroup 
  \setlength{\tabcolsep}{0.3pt} 
\fontsize{8.2}{13}\selectfont 
     \begin{tabular}{cccccccccccccccc}
    \toprule
    CR    & \multicolumn{3}{c}{4    } & \multicolumn{3}{c}{8    } & \multicolumn{3}{c}{16    } & \multicolumn{3}{c}{32    } & \multicolumn{3}{c}{64} \\
    \midrule
    \multirow{2}[2]{*}{Methods} & \multirow{2}[2]{*}{FLOPs} & \multicolumn{2}{c}{NMSE} & \multirow{2}[2]{*}{FLOPs} & \multicolumn{2}{c}{NMSE} & \multirow{2}[2]{*}{FLOPs} & \multicolumn{2}{c}{NMSE} & \multirow{2}[2]{*}{FLOPs} & \multicolumn{2}{c}{NMSE} & \multirow{2}[2]{*}{FLOPs} & \multicolumn{2}{c}{NMSE} \\
          &       & Indoor & Outdoor &       & Indoor & Outdoor &       & Indoor & Outdoor &       & Indoor & Outdoor &       & Indoor & Outdoor \\
    \midrule
    CsiConformer & 22.86M & -31.54  & \textbf{-16.36}  &  21.81M & \textbf{-23.46}  & \textbf{-10.98}  & 21.28M  & \textbf{-19.33 } & {-7.29}  & 21.02M & \textbf{-14.58}  & \textbf{-5.01}  & 20.89M  & \textbf{-9.29}  & \textbf{-3.19}  \\
    CsiConformer+Q & 22.89M & -30.81  & \textit{-16.79}  & 21.82M & \textit{-24.05}  & \textit{-11.21}  & 21.29M   &\textit{ -19.57}  & \textit{-7.40}  & 21.03M & -14.52  & -4.92  & 20.89M & \textit{-10.29}  & \textit{-3.44 } \\
    \cdashline{1-16}
    DCRNet & 17.57M & -30.61  & -13.72  & 16.52M & -19.92  & -10.17  & 16.00M & -14.02  & -6.35  & 15.74M & -9.88  & -3.95  & /     & /     & / \\
    TransNet & 35.72M  & \textbf{-32.38}  & -14.86  & 34.70M  & -22.91  & -9.99  & 34.14M  & -15.00  & \textbf{-7.82}  & 33.88M & -10.49  & -4.13  & 33.75M & -6.08  & -2.62  \\
    CsiNet+ & 24.57M & -27.37  & -12.40  & 23.52M & -18.29  & -8.72  & 23.00M  & -14.14  & -5.73  & 22.74M  & -10.43  & -3.40  & /     & /     & / \\
    \cdashline{1-16}
    CLNet & 4.05M & -29.16  & -12.88  & 3.01M & -15.60  & -8.29  & 2.48M & -11.15  & -5.56  & 2.22M & -8.95  & -3.49  & 2.09M & -6.34  & -2.19  \\
    CRNet & 5.12M & -24.10  & -12.57  & 4.07M  & -15.04  & -7.94  & 3.55M & -10.52  & -5.36  & 3.29M & -8.90  & -3.16  & 3.16M & -6.23  & -2.19  \\
      \bottomrule
    \end{tabular}
  \endgroup
  \label{tab1}%
  \begin{tablenotes}   
        \footnotesize            
        \item[1] / means the result is not reported in the original paper \cite{csinetjia,dilated}.
        \item[2] The best results in all networks without quantization are shown in bold. 
        \item[3] Results in italic  means that CsiConformer with SVQ-VAE outperforms CsiConformer without quantization in the same CR and scenario.
 \end{tablenotes}        
\end{threeparttable}  
\vspace{-20pt}
\end{table*}%
The total of 150,000 independently generated CSI are split into three parts, i.e., 100,000 for training, 30,000 for validation, and 20,000 for testing with batches of 200 samples each. Cosine annealing scheme is adopted with warm-up epochs set to 30 and the learning rate ranges between  \(5\times10^{-5}\) and \(2\times10^{-4}\). The proposed network CsiConformer is trained for 1000 epochs with Adam optimizer. The normalized mean square error (NMSE) is adopted to measure CSI reconstruction accuracy.

\subsection{CsiConformer and SVQ-VAE Overall Performance}
Table I shows the performance comparison between CsiConformer and previous SOTA networks including CsiNet+ \cite{csinetjia}, CRNet \cite{CRNet}, CLNet \cite{CLnet}, TransNet \cite{transnet}, and DCRNet \cite{dilated}  with different CRs. The computational overhead is measured using floating point operations (FLOPs) while the CSI feedback accuracy is evaluated by NMSE. 

As for complexity, CsiConformer requires 37.36\% fewer FLOPs on average compared with the SOTA TransNet, making it more suitable for UE deployment. Regarding CSI feedback accuracy, CsiConformer outperforms both lightweight and heavyweight networks across most CRs in indoor and outdoor scenarios except for the indoor scenario with CR \(=\!4\) and the outdoor scenario with CR \(=\!16\). The NMSE performance of  CsiConformer gains 17.67\% 
 (1.48dB) overall average  improvement compared with the SOTA TransNet, improved for 24.10\% and 11.24\% in indoor and outdoor scenarios, respectively. Noteworthy is CsiConformer's remarkable performance with high CRs. Compared to the best results of previous SOTA methods with CR \(=\!32\) and 64, CsiConformer achieves NMSE improvements  of 33.70\%  overall average. This significant  enhancement is mainly attributed to the efficient capture of relative-offset-based local correlations within the CSI matrix by the convolution modules in CsiConformer. Additionally, two half-step FCC modules in CsiConformer outperform the single FCC module in TransNet. Moreover, with 4 layers and a smaller multiple of change in the FCC module, CsiConformer surpasses TransNet's 2 layers with a larger expansion factor. Since CsiConformer evidently enhances CSI reconstruction accuracy, the additional computational overhead compared with lighter networks is manageable.

Furthermore, CsiConformer+SVQ-VAE scheme shows its performance in CsiConformer+Q of Table I. The quantizer introduces 1024 learnable parameters from a codebook size of \(32\times32\) embedding space. Regarding  accuracy, Conformer+SVQ-VAE occasionally outperforms the original CsiConformer without quantization when the number of quantization bits is 5. NMSE after quantization of SVQ-VAE is improved by 2.35\% and 2.47\% on average in indoor and outdoor scenarios, respectively. The reason is that SVQ-VAE  captures detailed  latent  features  after the up-channel convolution layer  increases the channel dimension of codewords, and quantization scheme  eliminates redundant information particularly in outdoor scenarios.
\subsection{Quantization Module Evaluation}
This subsection compares the proposed quantization module SVQ-VAE with the following quantization schemes: uniform quantization, \(\mu\)-law quantization, VQ-VAE-based quantization in \cite{rizzello2022learning}. Note that the above quantization methods should be compared under the same training mode.  The results of the retraining decoder training mode are always different from those of an end-to-end fashion, with the latter generally performs better \cite{clusterQUA}. Hence, for the uniform quantization and µ-law quantization, we adopt an end-to-end fashion and set the gradient of quantizers to constant one to pass the back propagation, similar to the approach used in  \cite{csinetjia,clusterQUA,changeable_rate_and_quafunction}. The training strategies and random seed are the same for all these quantization methods with 1000 epochs and cosine annealing learning rate. Table II illustrates the NMSE performance of CsiConformer with different quantization schemes. 
 \begin{table}[htbp]
  \centering
   \captionsetup{justification=centering}
   \caption{\centering{\scshape{NMSE (dB) Performance of CsiConformer with SVQ-VAE. Schemes Being Compared: Uniform Constant Gradient Quantizer(UQ-E2E); µ-Law Constant Gradient Quantizer (\(\mu\)Q-E2E); VQ-VAE-based quantizer in [18](base-VV).}}}
    \begingroup 
  \setlength{\tabcolsep}{5.5pt} 
 \renewcommand{\arraystretch}{1.0} 
  \begin{center}
\fontsize{8.2}{8}\selectfont 
    \begin{tabular}{c|c|c|ccc}
\cline{1-6} 
 \multirow{13}[7]{*}{Indoor} & Method & \diagbox[width=7em]{CR}{B} & 3 & 4 & 5 \bigstrut[b]\\

\cline{2-6}          & UQ-E2E    & \multirow{4}[2]{*}{4} & -11.84  & -18.12  & -22.56  \bigstrut[t]\\
          & \(\mu\)Q-E2E   &       & -16.90  & -20.93  & -24.20  \\
          & base-VV &       & -19.41  & -24.22  & -26.71  \\
          & SVQ-VAE &       & -22.97  & -27.11  & -30.81  \bigstrut[b]\\
\cline{2-6}          & UQ-E2E    & \multirow{4}[2]{*}{16} & -7.35  & -12.03  & -16.43  \bigstrut[t]\\
          & \(\mu\)Q-E2E   &       & -11.76  & -15.30  & -17.02  \\
          & base-VV &       & -8.36  & -16.34  & -18.46  \\
          & SVQ-VAE &       & -16.02  & -18.38  & -19.57  \bigstrut[b]\\
\cline{2-6}          & UQ-E2E    & \multirow{4}[2]{*}{64} & -6.52  & -7.87  & -8.62  \bigstrut[t]\\
          & \(\mu\)Q-E2E   &   & 1.23e-4      &  1.22e-4  & -8.63  \\
          & base-VV &       & -7.84  & -8.44  & -8.73  \\
          & SVQ-VAE &       & -8.20  & -9.70  & -10.29  \bigstrut[b]\\
\cline{1-6} 
    \multirow{12}[6]{*}{Outdoor} & UQ-E2E    & \multirow{4}[2]{*}{4} & -7.72  & -10.24  & -11.76  \bigstrut[t]\\
          & \(\mu\)Q-E2E   &       & -8.04  & -9.82  & -15.14  \\
          & base-VV &       & -11.31  & -13.20  & -14.82  \\
          & SVQ-VAE &       & -14.22  & -16.17  & -16.79  \bigstrut[b]\\
\cline{2-6}          & UQ-E2E    & \multirow{4}[2]{*}{16} & -3.21  & -5.80  & -6.52  \bigstrut[t]\\
          & \(\mu\)Q-E2E   &       & -4.62  & -6.27  & -6.87  \\
          & base\_VQ-VAE &       & -5.64  & -6.27  & -7.13  \\
          & SVQ-VAE &       & -6.69  & -6.91  & -7.40  \bigstrut[b]\\
\cline{2-6}          & UQ-E2E    & \multirow{4}[2]{*}{64} & -1.69  & -2.18  & -2.13  \bigstrut[t]\\
          & \(\mu\)Q-E2E   &       & -2.11  & -2.84  & -2.80  \\
          & base-VV &       & -2.25  & -2.40  & -3.13  \\
          & SVQ-VAE &       & -2.93  & -3.01  & -3.44  \bigstrut[b]\\
\cline{1-6} 
    \end{tabular}%
       \end{center}
      \endgroup
  \label{tab:addlabel}%
  \vspace{-5pt}
\end{table}%
Experiment results in Table II indicate that SVQ-VAE is obviously superior with improvement on NMSE compared with other quantization schemes. Traditional quantization schemes tend to be too coarse in matching the distribution of CSI codeword elements, especially with lower quantization bits. The proposed SVQ-VAE  better matches the distribution of codewords using an up-channel convolution and a codebook with long latent embedding vectors. 
\subsection{Ablation Studies}
This subsection examines the influence of the convolution module and the multi-layer structure in the CSI feedback network. The  settings remain consistent with those described above. CsiConformerII modifies the number of CinConformer layers and the expansion factor of the FCC module to 3 and 6,  respectively, ensuring a similar computational overhead. The baseline has FLOPs of 21.28M and 21.02M with CRs of 16 and 32, while CsiConformerII has FLOPs of 22.39M and 22.12M with the same CRs,  respectively.

As shown in Table III, our model experiences performance degradation without the convolution module, and increasing the number of layers instead of the expansion factor with nearly the same computational complexity leads to  higher accuracy in reconstruction.
\begin{table}[htbp]
  \centering
   \captionsetup{justification=centering}
  \caption{ NMSE (dB) Performance of CsiConformer With Different Structures}
  \fontsize{8.2}{7}\selectfont 
    \begin{tabular}{|c|c|c|c|c|c|c|}
\cline{1-7} 
    \multirow{2}[4]{*}{  CR  } & \multicolumn{2}{c|}{Baseline} & \multicolumn{2}{c|}{None\_cov}  & \multicolumn{2}{c|}{CsiConformerII} \bigstrut\\
\cline{2-7}          & Indoor & Outdoor & Indoor & Outdoor  & Indoor & Outdoor \bigstrut\\
\cline{1-7} 
    16     & -19.33 & -7.29 & -17.78 & -6.80 &-16.82 &-6.86 \bigstrut\\
\cline{1-7} 
    32   & -14.58 & -5.01  & -12.54 & -4.71 & -13.85 & -4.79 \bigstrut\\
\cline{1-7} 
    \end{tabular}%
  \label{tab:addlabel}%
  
\begin{tablenotes}[flushleft]
    \footnotesize            
    \item  \(^{\small1}\) Baseline denotes the results of CsiConformer from Table I.
    \item  \(^{\small2}\) None\_conv removes the CNN module entirely from CsiConformer.
\end{tablenotes} 
\vspace{-15pt}
\end{table}%
\section{Conclusion}
This letter presents a novel DL-based model named CsiConformer for CSI  reconstruction in massive MIMO FDD systems. Compared with  previous SOTA networks, CsiConformer achieves obvious performance improvement in CSI reconstruction quality with lower FLOPs. To further improve the encoding efficiency, we also propose a new quantization module named SVQ-VAE. The integrated  CsiConformer+SVQ-VAE structure achieves superior CSI feedback quality and efficiency. 
\bibliography{text}

\begin{thebibliography}{10}
\providecommand{\url}[1]{#1}
\csname url@samestyle\endcsname
\providecommand{\newblock}{\relax}
\providecommand{\bibinfo}[2]{#2}
\providecommand{\BIBentrySTDinterwordspacing}{\spaceskip=0pt\relax}
\providecommand{\BIBentryALTinterwordstretchfactor}{4}
\providecommand{\BIBentryALTinterwordspacing}{\spaceskip=\fontdimen2\font plus
\BIBentryALTinterwordstretchfactor\fontdimen3\font minus \fontdimen4\font\relax}
\providecommand{\BIBforeignlanguage}[2]{{%
\expandafter\ifx\csname l@#1\endcsname\relax
\typeout{** WARNING: IEEEtran.bst: No hyphenation pattern has been}%
\typeout{** loaded for the language `#1'. Using the pattern for}%
\typeout{** the default language instead.}%
\else
\language=\csname l@#1\endcsname
\fi
#2}}
\providecommand{\BIBdecl}{\relax}
\BIBdecl

\bibitem{hong2017multibeam}
W.~Hong, Z.~H. Jiang, C.~Yu, J.~Zhou, P.~Chen, Z.~Yu, H.~Zhang, B.~Yang, X.~Pang, M.~Jiang \emph{et~al.}, ``Multibeam antenna technologies for 5g wireless communications,'' \emph{IEEE Transactions on Antennas and Propagation}, vol.~65, no.~12, pp. 6231--6249, 2017.

\bibitem{sanguinetti2019toward}
L.~Sanguinetti, E.~Bj{\"o}rnson, and J.~Hoydis, ``Toward massive mimo 2.0: Understanding spatial correlation, interference suppression, and pilot contamination,'' \emph{IEEE Transactions on Communications}, vol.~68, no.~1, pp. 232--257, 2019.

\bibitem{kuo2012compressive}
P.-H. Kuo, H.~Kung, and P.-A. Ting, ``Compressive sensing based channel feedback protocols for spatially-correlated massive antenna arrays,'' in \emph{2012 IEEE Wireless Communications and Networking Conference (WCNC)}.\hskip 1em plus 0.5em minus 0.4em\relax IEEE, 2012, pp. 492--497.

\bibitem{kyritsi2003correlation}
P.~Kyritsi, D.~C. Cox, R.~A. Valenzuela, and P.~W. Wolniansky, ``Correlation analysis based on mimo channel measurements in an indoor environment,'' \emph{IEEE Journal on Selected areas in communications}, vol.~21, no.~5, pp. 713--720, 2003.

\bibitem{wen2018deep}
C.-K. Wen, W.-T. Shih, and S.~Jin, ``Deep learning for massive mimo csi feedback,'' \emph{IEEE Wireless Communications Letters}, vol.~7, no.~5, pp. 748--751, 2018.

\bibitem{liu2019exploiting}
Z.~Liu, L.~Zhang, and Z.~Ding, ``Exploiting bi-directional channel reciprocity in deep learning for low rate massive mimo csi feedback,'' \emph{IEEE Wireless Communications Letters}, vol.~8, no.~3, pp. 889--892, 2019.

\bibitem{wang2018deep}
T.~Wang, C.-K. Wen, S.~Jin, and G.~Y. Li, ``Deep learning-based csi feedback approach for time-varying massive mimo channels,'' \emph{IEEE Wireless Communications Letters}, vol.~8, no.~2, pp. 416--419, 2018.

\bibitem{9839111}
X.~Liang, J.~Shen, H.~Chang, X.~Gu, and L.~Zhang, ``Deep learning-based cooperative csi feedback via multiple receiving antennas in massive mimo,'' in \emph{ICC 2022 - IEEE International Conference on Communications}, 2022, pp. 1373--1378.

\bibitem{csinetjia}
J.~Guo, C.-K. Wen, S.~Jin, and G.~Y. Li, ``Convolutional neural network-based multiple-rate compressive sensing for massive mimo csi feedback: Design, simulation, and analysis,'' \emph{IEEE Transactions on Wireless Communications}, vol.~19, no.~4, pp. 2827--2840, 2020.

\bibitem{CRNet}
Z.~Lu, J.~Wang, and J.~Song, ``Multi-resolution csi feedback with deep learning in massive mimo system,'' in \emph{ICC 2020-2020 IEEE international conference on communications (ICC)}.\hskip 1em plus 0.5em minus 0.4em\relax IEEE, 2020, pp. 1--6.

\bibitem{CLnet}
S.~Ji and M.~Li, ``Clnet: Complex input lightweight neural network designed for massive mimo csi feedback,'' \emph{IEEE Wireless Communications Letters}, vol.~10, no.~10, pp. 2318--2322, 2021.

\bibitem{transnet}
Y.~Cui, A.~Guo, and C.~Song, ``Transnet: Full attention network for csi feedback in fdd massive mimo system,'' \emph{IEEE Wireless Communications Letters}, vol.~11, no.~5, pp. 903--907, 2022.

\bibitem{dilated}
S.~Tang, J.~Xia, L.~Fan, X.~Lei, W.~Xu, and A.~Nallanathan, ``Dilated convolution based csi feedback compression for massive mimo systems,'' \emph{IEEE Transactions on Vehicular Technology}, vol.~71, no.~10, pp. 11\,216--11\,221, 2022.

\bibitem{hu2023learnable}
Z.~Hu, G.~Liu, Q.~Xie, J.~Xue, D.~Meng, and D.~G{\"u}nd{\"u}z, ``A learnable optimization and regularization approach to massive mimo csi feedback,'' \emph{IEEE Transactions on Wireless Communications}, 2023.

\bibitem{clusterQUA}
J.~Shen, X.~Liang, X.~Gu, and L.~Zhang, ``Clustering algorithm-based quantization method for massive mimo csi feedback,'' \emph{IEEE Wireless Communications Letters}, vol.~11, no.~10, pp. 2155--2159, 2022.

\bibitem{changeable_rate_and_quafunction}
X.~Liang, H.~Chang, H.~Li, X.~Gu, and L.~Zhang, ``Changeable rate and novel quantization for csi feedback based on deep learning,'' \emph{IEEE Transactions on Wireless Communications}, vol.~21, no.~12, pp. 10\,100--10\,114, 2022.

\bibitem{conformer}
A.~Gulati, J.~Qin, C.-C. Chiu, N.~Parmar, Y.~Zhang, J.~Yu, W.~Han, S.~Wang, Z.~Zhang, Y.~Wu \emph{et~al.}, ``Conformer: Convolution-augmented transformer for speech recognition,'' \emph{arXiv preprint arXiv:2005.08100}, 2020.

\bibitem{vqvae}
A.~Van Den~Oord, O.~Vinyals \emph{et~al.}, ``Neural discrete representation learning,'' \emph{Advances in neural information processing systems}, vol.~30, 2017.

\bibitem{rizzello2022learning}
V.~Rizzello, M.~Nerini, M.~Joham, B.~Clerckx, and W.~Utschick, ``Learning representations for csi adaptive quantization and feedback,'' \emph{arXiv preprint arXiv:2207.06924}, 2022.

\bibitem{cost2100}
L.~Liu, C.~Oestges, J.~Poutanen, K.~Haneda, P.~Vainikainen, F.~Quitin, F.~Tufvesson, and P.~De~Doncker, ``The cost 2100 mimo channel model,'' \emph{IEEE Wireless Communications}, vol.~19, no.~6, pp. 92--99, 2012.

\end{thebibliography}
\bibliographystyle{IEEEtran}

\end{document}